\documentclass{aa}
\usepackage{graphicx,txfonts,natbib,color}
\bibpunct{(}{)}{;}{a}{}{,}

\begin{document}
\title{Influence of the circumbinary disk gravity on planetesimal 
accumulation in the Kepler--16 system}
\author{F. Marzari\inst{1}, P. Thebault \inst{2}, H. Scholl
\inst{3}, G. Picogna \inst{1}
and C. Baruteau \inst{4} }
\institute{
  Dipartimento di Fisica, University of Padova, Via Marzolo 8,
  35131 Padova, Italy 
  \and
  Observatoire de Paris, Section de
  Meudon,
  F-92195 Meudon Principal Cedex, France
  \and 
  Observatoire de la C{\^o}te d'Azur, B.P. 4229, F-06304 Nice Cedex, France
  \and
  DAMTP, University of Cambridge, Wilberforce Road, Cambridge CB30WA, United Kingdom
}
\titlerunning{Planetesimal accumulation in 16 Kepler B}
\authorrunning{F. Marzari et al.}
\abstract 
{Recent observations from NASA's Kepler mission detected 
the first planets in circumbinary orbits. The question we
try to answer is where these planets formed in the 
circumbinary disk and how far inside they migrated to 
reach their present location. 
}
{We investigate the first and more delicate phase of 
planet formation when planetesimals accumulate to
form planetary embryos.  
}
{ We use the hydrodynamical code FARGO to study the 
evolution of the disk and of a test population of 
planetesimals embedded in it. With this hybrid 
hydrodynamical--N--body code we
can properly account for the gas drag force on the 
planetesimals and for the gravitational force of 
the disk on them. 
}
{The numerical simulations show that the gravity
of the eccentric disk on the planetesimal swarm 
excites their eccentricities to values much larger than those 
induced by the binary perturbations only 
within 10 AU from the stars. Moreover, the disk 
gravity
prevents a full alignment of the planetesimal
pericenters. Both these effects
lead to large impact velocities, 
beyond the critical value for erosion. 
}
{Planetesimals accumulation in circumbinary disks appears 
to be prevented close to the stellar pair by the 
gravitational perturbations of the circumbinary disk. 
The observed planets possibly formed in the outer regions 
of the disk and then migrated inside by tidal interaction 
with the disk. 
}
\maketitle
\section{Introduction} 
\label{intro}

Most Sun-like stars are believed to form as gravitationally bound pairs.
Estimating the prevalence of planets in binary systems is then 
a relevant issue when computing the fraction of stars with planets. 
Planets in circumstellar orbits (S--type orbits) around one of
the components of a binary system
have been found in both close ($\sim$20\,AU) and wide binary systems confirming that 
the binary perturbations may not be able to prevent planet formation 
even if they can significantly affect the course of it. This last
issue has been explored in detail using numerical simulations, most of them 
modeling the intermediate stage of planet formation, the accumulation of 
kilometer-sized planetesimal 
\citep{mascho,thebs04,theb06,thems08,klene08,
thems09,mabash,xie09,xie10,paard08,marba2}. 
These studies have shown that this phase might be the most sensitive
to binarity effects, because mutual impact velocities can be increased
to values that may threaten the formation of large bodies.
Crucial in this phase is the evolution
of both the eccentricity and perihelion longitude of the planetesimals
under the coupled action of the companion gravity and gas drag force. 
A size-dependent phasing of the orbits develops over the timescale
of the secular perturbations of the binary, leading to high (and
accretion inhibiting) collision
velocities among any planetesimal population with even a small 
size spread \citep{theb06}.
Mechanisms that could come to the rescue of the planetesimal 
accumulation process 
are a small inclination between the circumstellar disc 
and the binary plane \citep{xie10}, 
or an outward migration of protoplanetary embryos formed in safer 
regions closer to the
star \citep{Payne09}, or planetesimal growth 
through the sweeping of small collisional fragments
\citep{Parli10,xie10b}, or the fact that the 
binary was initially wider and was compacted
by stellar encounters during the early evolution of the stellar 
cluster it was born in \citep{theb09}.
Another, more radical solution would be that planet 
formation proceeds through a different channel
in close binaries, a hypothesis that could be supported by the fact 
the exoplanets found
in close binaries have different properties than those around 
single stars \citep{Duchene10}.
For a more detailed discussion on circumprimary planet formation in binaries, see the recent review by \citet{theb11}.

Recently, the KEPLER team announced the discovery of another category of exoplanets in binaries,
 i.e, transiting circumbinary planets, which are planets moving on P--type orbits circumventing both the 
stars. The first of such planets to be discovered was Kepler--16 b
\citep{doyle11} followed by Kepler-34 b,  Kepler-35 b \citep{welsh12}
and Kepler-47 b,c, the first circumbinary multi--planet system
\citep{orosz12}. 
The formation of planets in circumbinary 
P--type orbits is very different compared to that in S--type orbit (around a
component of the pair) since the secular perturbations of the companion 
star on a planetesimal swarm 
have different intensity and form compared to that produced 
by an external perturber. As a consequence, the outcomes of numerical 
modeling of planetesimal evolution in S--type orbits cannot be applied 
to study planet formation in circumbinary orbits. 

At present, two numerical studies have been performed to estimate 
where in the initial circumbinary disk around Kepler--16
planetesimal accumulation can proceed, even if perturbed 
and possibly at a slower pace, towards the accumulation of 
large planetary embryos. Both studies \citep{meschia12,paar12} used 
an N--body approach to compute the trajectories of a large number of 
planetesimals perturbed by the non--spherically symmetric gravitational 
field of the stellar pair. The first paper \citep{meschia12} focused 
on the long term effects of the secular perturbations finding 
potential accretion-friendly zones within 1.75 AU from the star pair 
and beyond 4 AU. In these regions the mutual impact 
velocity between planetesimals were, in the majority of collisions, 
lower than the threshold velocity causing disruption. Their
eccentricity distribution is centered around the 
the forced component of the 
secular perturbations of the binary given by 

\begin{equation}
e_f = {{5} \over {4}} (1 -2 \mu)  {{a_B} \over {a}} e_B,
\end{equation}

where $\mu = m_2/(m_1+m_2)$ is the 
binary mass ratio, $a_B$ and $e_B$ the semimajor
axis and eccentricity of the binary system and $a$ is the semimajor axis of 
the planetesimal \citep{morna04}. 

In a subsequent paper, \cite{paar12}  re-analysed 
planetesimal accumulation around Kepler--16 and also explored 
two other systems, Kepler--34 and Kepler--35. They focused on the 
effects of short term eccentricity perturbations on the 
planetesimal motion. 
They also  considered 
the possible reaccumulation of small fragments, produced by the shattering
of planetesimals impacting at high relative velocity, 
onto the largest intact planetesimals.
Even if this second-generation accretion helps, \citet{paar12} conclude that 
planet formation is not possible at the present location of the 
planet, but may however be effective beyond twice the present 
semimajor axis ($a = 0.7$ AU). 
Both the previously mentioned studies \citep{meschia12,paar12}
claim that, in spite of the combined perturbations of the binary star system and 
of the friction
from the gas of the circumbinary disk, planet formation seems to be 
possible beyond a distance of about 2--4 AU from the baricenter of the pair,
which is further out than the present location of the exoplanets but still relatively
close to the binary.

The basic assumption of an N-body approach to the problem 
of planetesimal accumulation requires that the gas disk in which 
the planetesimals are embedded is axisymmetric. This approximation 
may not be a good one when dealing with circumbinary disks 
since the gravitational perturbations of the star pair   
may strongly affect the disk shape. As already shown in 
\citep{kle08,marba1,marba2,kle12}, the disk may become eccentric and be 
perturbed by strong spiral waves. In that paper, 
the parameters of the system were similar to those of Kepler--16 but
the disk was truncated at 3 AU and the isothermal 
approximation was used.  In this paper we adopt a more complete approach 
modeling the circumbinary disk that gave origin to the planet in 
Kepler--16 as a radiative disk extending out to 10 AU from the 
baricenter of the two stars. The radiative model is better suited to 
describe the earlier stages of the disk evolution when it is massive
and probably optically thick. It has also been 
shown \citep{marba2,kle12} that radiative disks,
when perturbed, develop 
overall shapes and internal structure that
differ from those of locally isothermal 
disks, in particular concerning
the disk eccentricity and the propagation of spiral waves
which may significantly perturb planetesimal trajectories.
For this reason, in order to obtain an accurate 
modeling of the disk structure, in our simulations we solve 
an energy equation that includes
the viscous heating of the disk and radiative losses. 
On the basis on the above mentioned references we 
expect that a detailed treatment of
the disk thermodynamics is more relevant for the 
evolution of the disk eccentricity
than the 
inclusion of the effects of self--gravity.
In addition, considering a larger disk allows a better 
handling of the gravitational perturbations of the disk 
on the planetesimal orbits. If the disk is more massive,
since we model a larger portion of it, the 
perturbations due to a potentially uneven mass distribution, due to the 
building up of an eccentric shape, are considerably stronger. 

In this paper we compute the trajectories of a large number of 
planetesimals perturbed by the primordial circumbinary disk. 
Our simulations focus on the Kepler--16 system and show that the 
gravitational perturbations of the disk excite large eccentricity 
values in the planetesimal swarm with a mechanism similar to that 
described by \cite{grenel10}. These large eccentricities,
up to 0.4 and beyond,
are able to prevent the onset of accumulation within 10 AU from 
the stars and maybe also farther out. The present planet observed 
in the system possibly formed in the outer regions of the 
disk and migrated inside as described in \cite{pinel07,pinel08}. 
The migration possibly occurred in all the circumbinary systems 
mentioned above and is strongly suggested by the coincidence 
between the semimajor axis of the planets and the  location of 
the internal stability limit which can be derived from \cite{holwi}.
As a consequence, our results do not suggest that 
planet formation is not possible, but that it had to occur in
the outer regions of the disk where the perturbations of the 
star pair are less effective in producing non-axisymmetric density
perturbations on the disk.  

In Section 2 we describe the numerical model and in Section 3
we describe out results. Section 4 is devoted to the discussion of
the results. 

\section{The hybrid algorithm modeling the evolution
of the disk and planetesimals} 
\label{model}

To compute the trajectories of planetesimals and, at the same time, 
the evolution of the gaseous disk under the perturbations of 
the binary, we have used the two-dimensional 
numerical code FARGO \citep{fargo2} modified to fit the problem. 
The hydrodynamical equations
are solved in a cylindrical coordinate system centered
on the baricenter of the binary. We focus on the Kepler--16 system
where the mass of the primary is $M_1 = 0.69$ $M_{\odot}$, that of 
the secondary $M_2 = 0.20$ $M_{\odot}$, the binary semimajor axis 
$a_B = 0.224$ $AU$ and the eccentricity $e_B = 0.159$.
The two stars are evolved on a 
fixed Keplerian orbit neglecting  
any change in the binary system due to gas accretion 
by the stars and any momentum exchange with the disk.
This choice is justified by the models of \cite{pinel07} 
suggesting that a system made of the binary and the disk reach 
a near-stationary state after some evolution.
Incidentally, these authors used binary parameters actually close
to those derived for Kepler--16. 

The grid used in our calculations to model the disk has $N_{\rm
  r} = 256$ radial zones and $N_{\rm s} = 512$ azimuthal zones, and an
arithmetic spacing is used along the radial direction
i.e. the radial distance is divided in equal 
size intervals. All the 
simulations are carried out including an energy equation of the 
form \citep{barmas08,marba2}

 \begin{equation}
    \frac{\partial e}{\partial t} + {\bf \nabla} \cdot (e {\bf v}) 
    = -p {\bf \nabla} \cdot {\bf v} 
    + Q^{+}_{\rm visc}
    - Q^{-}_{\rm cool}
    + \lambda e \nabla^2 \log(p/\Sigma^{\gamma})
  \label{eqnenergy}
  \end{equation}
where $e$ = $p / (\gamma-1)$ is the thermal energy density, $\gamma=1.4$
is the adiabatic index, and ${\bf v}$ denotes the gas velocity.  
In the equation we do not include 
the effects of stellar irradiation. 
The term 
$Q^{+}_{\rm visc}$ is the heating term due to the viscous heating
while  
the cooling term
 $Q^{-}_{\rm cool}$ is
assumed to be $2
  \sigma_{\rm SB} T_{\rm eff}^4$, where $\sigma_{\rm SB}$ is the
Stefan-Boltzmann constant and $T_{\rm eff}$ is the effective
temperature estimated as \citep{Hubeny90}
  \begin{equation}
  T^4_{\rm eff} = T^4 / \tau_{\rm eff},
  \end{equation}
for an effective optical depth
  \begin{equation}
  \tau_{\rm eff} = \frac{3\tau}{8} + \frac{\sqrt{3}}{4} + \frac{1}{4\tau}.
  \end{equation}
The vertical optical depth, $\tau$, is approximated as $\tau =
\kappa\Sigma / 2$, where for the Rosseland mean opacity, $\kappa$,
we adopt the formulae in \cite{beli}. Following \cite{pbk11}, we
also model thermal diffusion as the diffusion of the gas entropy, $s$,
defined as $s = {\cal R}(\gamma-1)^{-1} \log(p /
\Sigma^{\gamma})$. This corresponds to the last term in the
right-hand side of Eq.~(\ref{eqnenergy}), where $\lambda$ is a
constant thermal diffusion coefficient. Throughout this study, we
adopt $\lambda=10^{-6}$ in code units.
The initial aspect ratio $ h = H/r$ is constant all over
the disk and is set to 0.05. A constant shear kinematic viscosity
of $10^{-5}$ (normalized units),  which  corresponds at about 
5 AU within the disk
to an $\alpha$ value of about $2.5 \times 10^{-3}$ (\cite{shak}), 
is used and open boundary
conditions are adopted with standard outflow at 
both the inner and outer edge. 
The
initial temperature profile of the disk is computed as $T(r) = T_0 r^{-1}$
where the value at 1 AU is set to
$T_0 = 630\,{\rm K}$. This value is derived following the 
approach described in \cite{marba2} and it depends on the
initial choice of the aspect ratio $ h$. 
The Toomre-Q parameter 
is quite large in the inner
disk parts, where the binary's perturbation is the strongest.
Its radial dependence is approximately $ Q(R) \sim 100 \times
(R / 1 AU)^{-3/2}$ and a value of about 10 is measured at R = 5 AU.

We neglect in our model the apsidal precession of 
the binary due to its interaction with the disk. 
According to \cite{rafi13}, the period of the 
binary precession due to an axisymmetric disk is 
given by 

\begin{equation}
T_{\tilde\omega} = 8 \pi \left ( {{M_B} \over {M_d}} \right ) 
\left ( {{r_o^{1/2} r_{in}^{5/2}}
\over { {a_B}^3 \tilde\phi n_B}}  \right )
\label{pre}
\end{equation}

where $M_d$ is the disk mass, $M_b$ the summ of the star masses, 
$n_B$ the mean motion of the binary, $r_{in}$ the inner border 
of the disk and 
$r_o$ the outer one, and $\tilde\phi$ is a constant  whose value 
depends on the ratio of $a_B$ and $r_{in}$ and on the 
mass ratio and it can be approximated by 0.5.
Using this equation we find that  
$T_{\tilde\omega} \sim 3 \times 10^4$ yrs (after some initial evolution, 
the mass of the disk settles to $M_d = 2.75 \times 
10^{-2} M_{\odot}$). This is longer than the timescale over 
which the planetesimal eccentricity grows i.e. $10^3$ yrs.
We expect that the binary precession might have 
an effect on the long term evolution of the 
system but not on the 
short timespan we are covering with the model. 
In addition, since the disk in our simulations is eccentric,
Eq.~(\ref{pre}) may not be very precise being derived 
under the assumption of an axisymmetric disk. As 
a consequence, a full numerical approach is needed  
since the formula for the potential of a axisymmetric 
disk, used in \cite{rafi13}, cannot be applied. 
We plan to explore this 
effect in the future hoping in an increase in computing
power.

The forces acting on the planetesimals include the binary 
gravitational force, 
the gravitational force exerted by  
the disk  and the gas drag force. 
The latter is 
calculated in the Stokes regime as \citep{adachi76}

\begin{equation}
{\bf F} = k v {\bf v}
\end{equation}
 
where v is the relative velocity between the planetesimal and the gas 
and the parameter {\it k} (Kary et al. 1993) is given by

\begin{equation}
k = {{3 \rho_g C_D} \over {8 \rho_p R_p}} 
\end{equation}

where $C_{D}$ is the aerodynamic drag coefficient
for objects
with large Reynold's number,
$\rho _{\mathrm{p}\,}$ 
and $R_{\mathrm{p}\,}$ are planetesimal density and radius, 
respectively, and $\rho _{g} $ is the gas density in the midplane derived from 
the surface mass density $\Sigma ( r)$ through the relation 

\begin{equation}
\rho _{g}=\Sigma ( r) / [ ( 2\pi ) ^{1/ 2}H] 
\end{equation}

with $H$ the disk scale height 
\citep{gukle02}. 
Respect to \citep{marza09} we adopt a more
extended disk ranging from 0.5 to 10 AU from the binary system with a
superficial density profile $\Sigma = \Sigma_0 r^{-1/2}$ where 
$\Sigma_0$ is the density at 1 AU set to
$\Sigma_0 = 2.5 \times 10^{-4}$ in normalized units,
i.e. $\sim 2.2 \times  10^3$ $g/cm^2$, compatible with 
Minimum Mass Solar 
Nebula density. 

To refine the computation of the drag force acting on the planetesimals 
we extrapolate the density and velocity $(\rho, {\bf v})$ 
of the gas at the 
planetesimal location with a bilinear fit from the 
values at the borders of each grid cell. This is a refinement  of 
the algorithm used in \cite{marza09}. 

\section{The results}
\label{results}

In this section we compute the orbital evolution of the 
planetesimals and of the disk. Before including the planetesimals
in the model, we evolve the disk for $10^5$  
binary revolutions (approximately
$10^4$ yr).
We then restart the simulation including 400 planetesimals on 
initially circular orbits with semimajor axis equally spaced from 
1 to 8.8 AU and we compute their orbits and the disk evolution
for $10^4$ yrs.  

\subsection{The disk shape}

\begin{figure}[hpt]
\resizebox{90mm}{!}{\includegraphics[angle=-90]{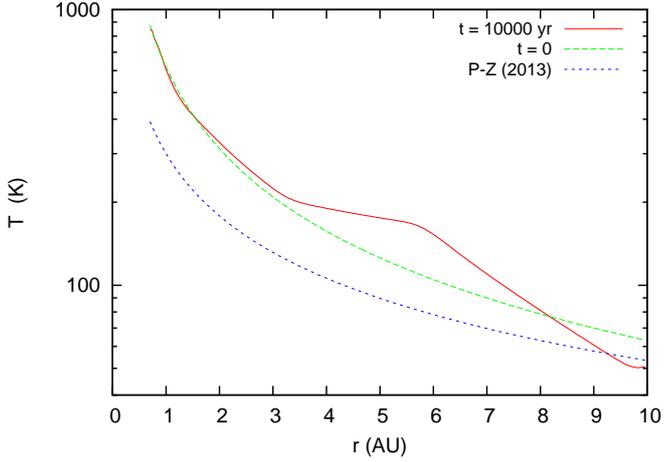}}
\caption{\label{ftemp} Azimuthally and time averaged (200 yr)
temperature profile of our radiative model at $t = 10000$ yr
(continuous red line) 
compared with the initial $t = 0$ yr non--equilibrium one
(dashed green line) and the isothermal temperature 
profile adopted by \cite{pelu13} (dotted blue line). 
     }
\end{figure}

\begin{figure}[hpt]
\resizebox{90mm}{!}{\includegraphics[angle=-90]{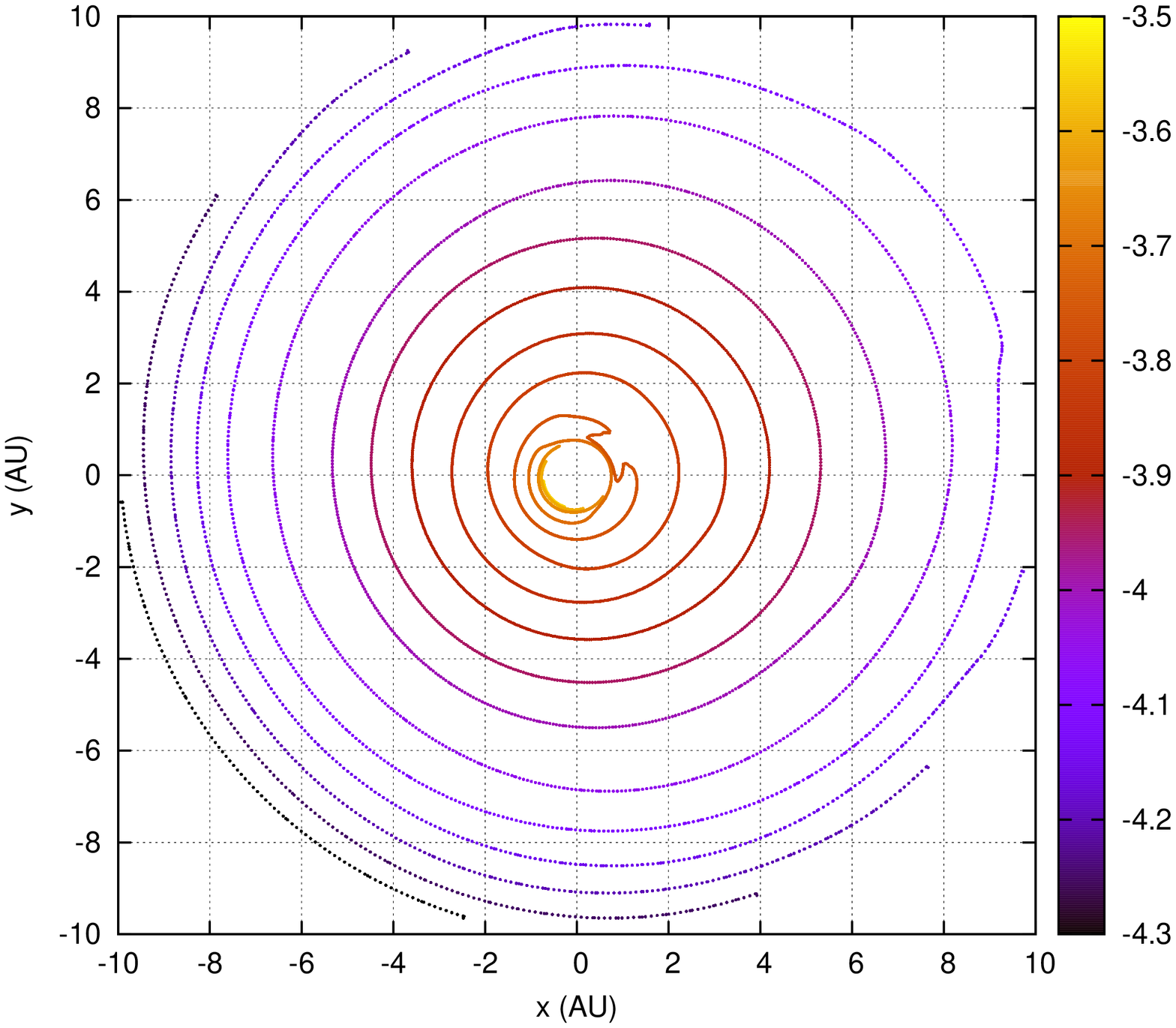}}
\resizebox{80mm}{!}{\includegraphics[angle=-90]{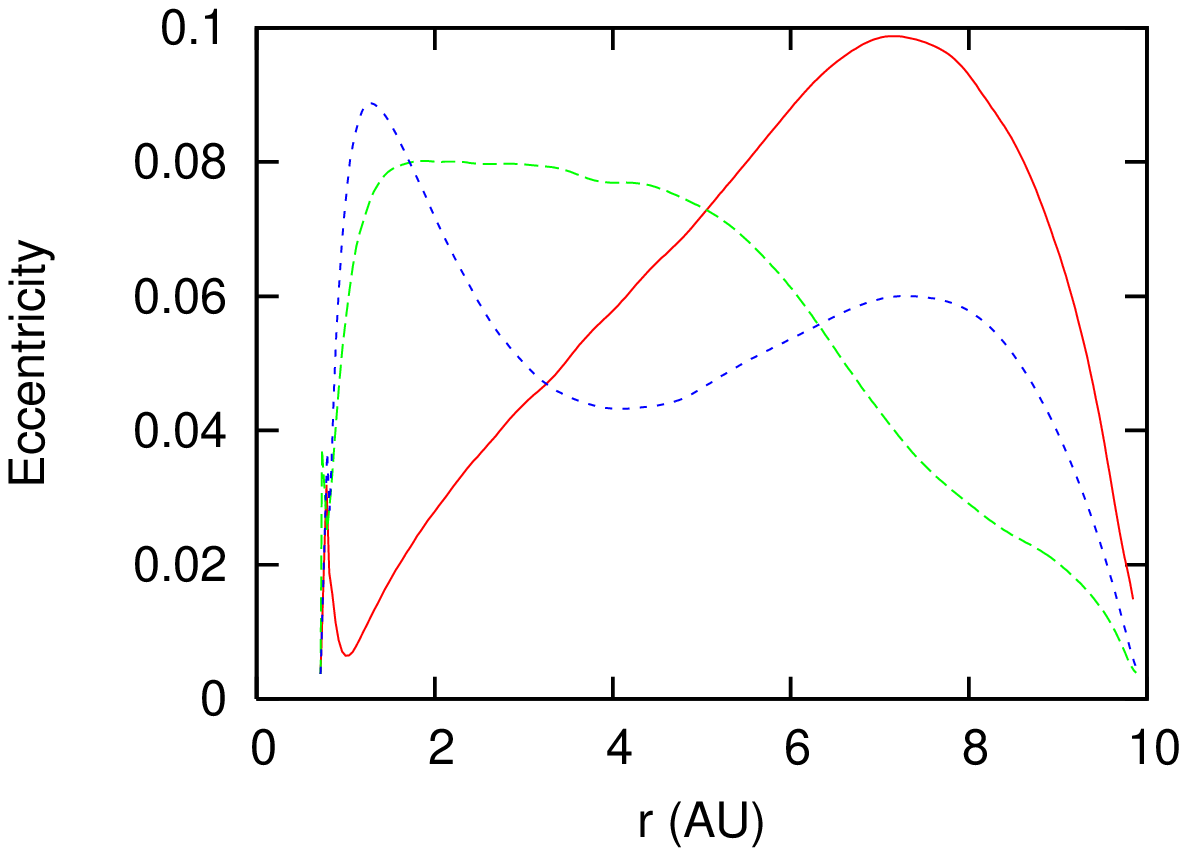}}
\caption{\label{fd1} 
In the upper panel we show isocurves of the 
disk density distribution (in normalized units) after 
$2 \times 10^5$ binary
revolutions  
illustrating the eccentric shape of the disk. In the bottom panel 
the azimuthally-averaged disk eccentricity is drawn as a
function of the radial distance at 3 different evolutionary times 
separated by 200 yr. 
     }
\end{figure}

In the upper panel of Fig.~\ref{fd1} we show isocurves of 
the surface density distribution of the disk confirming its
eccentric shape
responsible for strong gravitational
perturbations on the planetesimal orbits. In the bottom panel 
we show the azimuthally-averaged disk eccentricity as a function 
of the distance from the stars, computed as in \cite{marba2}, 
for 3 different evolutionary times separated by 200 yr. 
The 3 curves suggest a strong variability of the internal 
disk shape with time.  This is due to the 
propagation of density waves within the disk and was observed
also in \cite{marba2} for circumstellar disks in binaries. 
The behaviour shown in Fig.~\ref{fd1}, lower panel, differs 
from that illustrated in \cite{pelu13} but the two 
disk models are substantially different. \cite{pelu13}
adopt a locally isothermal equation of state with a fixed temperature 
profile $T(r) \sim 300 \times   R^{-3/4}$ while our initial temperature 
profile is $T(r) \sim 630 \times R^{-1}$ and then it evolves 
in time due to radiative cooling and various sources of heating, 
including that arising from shock waves. 
In Fig.~\ref{ftemp} we compare the temperature
profile of the isothermal disk model by \cite{pelu13} with 
the averaged (over 200 yr)  equilibrium profile of 
our radiative model. The difference in both absolute values 
and slope are noteworthy and this justifies the significant
differences in terms of disk eccentricity.  

It has been shown in 
\cite{marba2} that the temperature profile and 
the adopted energy equation have strong effects on 
the disk eccentricity in perturbed disks. A different 
thermodynamical model may lead to significant differences
in the disk evolution and in particular in the eccentricity
which may change by
more than a factor 10.  
The temperature profile of our radiative disk 
is significantly higher compared to that 
of \cite{pelu13} and this plays in favor of a higher disk 
eccentricity in our model. On the other hand,  \cite{marba2}
have shown that 
wave propagation through adiabatic
compressions and expansions may be more efficiently damped
in presence of radiative losses. This might cause a reduction 
of the disk eccentricity. All these physical phenomena that 
can lead to a different disk shape are more important thatn 
the disk self--gravity which would have less influnce on the
disk eccentricity than the thermodynamical model. 
In addition, self--gravity would not significantly affect
the planetesimal dynamics in the inner regions due to the 
large value of the Toomre parameter Q. 
The initial different temperature profile is not the only 
difference between our model and that of 
\cite{pelu13}, also the superficial density is difference 
as theirs declines as $r^{-1}$ while ours follows a
$r^{-1/2}$ law. 

In a real disk, not only viscous heating and radiative cooling
determine the disk’s thermal structure, but also 
stellar irradiation. Recently, \cite{bit13} have shown that 
stellar irradiation dominates in the outer regions of a disk while 
viscous heating rules close to the star  where
the structure of disks with and without stellar irra-
diation are similar.
They find via numerical 
modeling that only beyond approximately 8 AU from the star stellar 
irradiation begins to have some influence on the evolution 
of a disk. However, a 
significant flaring of the disk is observed only for massive disks
with $\Sigma_0 \sim 3000 g/cm^3$ and low viscosity ($\alpha \sim 0.001$). 
In this case the flaring appears consistent beyond 30 AU from the star. 
In our model, the circumbinary disk of 16 Kepler extends out to 10 AU
so it is expected not to be strongly influenced by stellar irradiation 
but to be dominated by the viscous evolution. 
In addition, in circumbinary disks density waves
excited by the tidal gravity field of the binary 
propagate within the disk.
In three dimensions these waves act like fundamental 
modes which correspond to large surface distorsions 
in the disk \citep{Lubow98}. \cite{Boley05} have also shown that 
shock waves in 3--D disks cause
sudden increases in the disk scale height, a phenomenon called 
hydraulic jump. If the spiral waves excited by the binary perturbations
are also shock waves, the hydraulic jumps in the vertical direction 
can shield the outer disc from stellar irradiation. Disk self-shadowing
due to spiral density waves may strongly reduce the relevance of 
stellar irradiation in circumbinary disks.

The relatively fast changes of the disk eccentricity and 
of the disk gravity field 
has an additional perturbative effect
on the planetesimals trajectories favouring eccentricity excitation.  
The orientation of the 
disk changes slowly with time on a timescale longer than $10^4$ yrs. 
However, the computation of the planetesimal orbits is very 
time consuming and the timespan of each model is limited by
the amount of CPU requirement. This is also the reason
why we neglect the self-gravity of the disk in our simulations. 
A model without planetesimals has however shown that self--gravity 
does not significantly affect the shape and time variability
of the circumbinary 
disk adopted in our simulations.

\subsection{Orbital evolution of 5 km size planetesimals}

The three major sources of perturbations on the planetesimal 
orbits are:

\begin{itemize}

\item - The secular and short term perturbations of the 
gravity field of the binary

\item - The gas drag force

\item - The disk gravitational force

\end{itemize}

The first kind of perturbation is accounted for also in pure N--body models
(neglecting the mutual gravitational attraction of planetesimals) 
and its effects are summarized by Eqs. 1 and 2. The second term is instead
handled very differently in a purely axisymmetric approach and our model. 
If the axisymmetric approximation is adopted, the gas velocity is assumed 
to point always in the tangential direction respect a circle centered 
in the binary baricenter and with radius equal to the planetesimal 
osculating radial 
distance. Its 
modulus is the local Keplerian velocity reduced by a factor that accounts 
for the pressure term $ v_g = v_K (1 - 2 \eta)^{1/2}$ where $\eta$ is of the 
order of $10^{-3}$. In our model, the gas velocity is computed directly 
from the solution of the hydrodynamical equations. This is an important 
aspect since the gas velocity, due to the disk elliptic shape and the 
presence of spiral waves, is very different from that computed in the 
axisymmetric approximation. The gas velocity is no longer circular 
and it can have a significant 
radial component. Its modulus can be much larger than the value estimated
by the previous simplified formula. In addition, the cylindrical symmetry 
is lost and the direction and modulus of the gas velocity depend on the 
azimuthal angle. 

The third perturbative component on the motion of planetesimals, that 
we will show to be the dominant one, is due to the gravity field of 
the disk. An asymmetric distribution of mass causes a significant 
perturbation of planetesimal trajectories. A similar phenomenon was
also observed  
by \cite{grenel10} 
in fully turbulent disks where embedded planetesimals 
develop large mutual encounter velocities due to
stochastic gravitational forces caused by turbulent density fluctuations.
In our scenario the density has a shaped asymmetric pattern and this 
is a worse source of perturbation since  its effect does not average 
to 0. In addition, the orientation of the perihelia play an important
role in our scenario \citep{theb06,mascho}, in particular when 
large eccentricities are excited. The 
gravitational perturbations of the eccentric disk significantly 
decrease 
the level of perihelia alignment of same size planetesimals
and this has important consequences on the accretion process. 

\begin{figure}[hpt]
\resizebox{90mm}{!}{\includegraphics[angle=0]{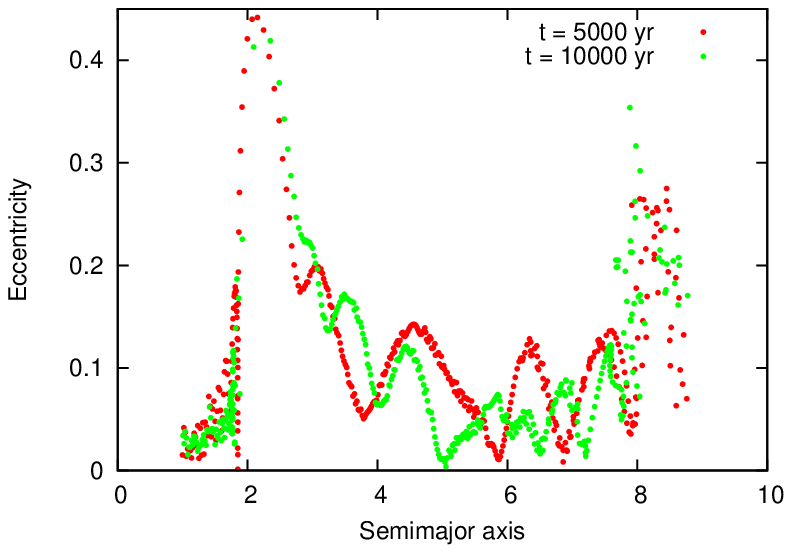}}
\resizebox{90mm}{!}{\includegraphics[angle=0]{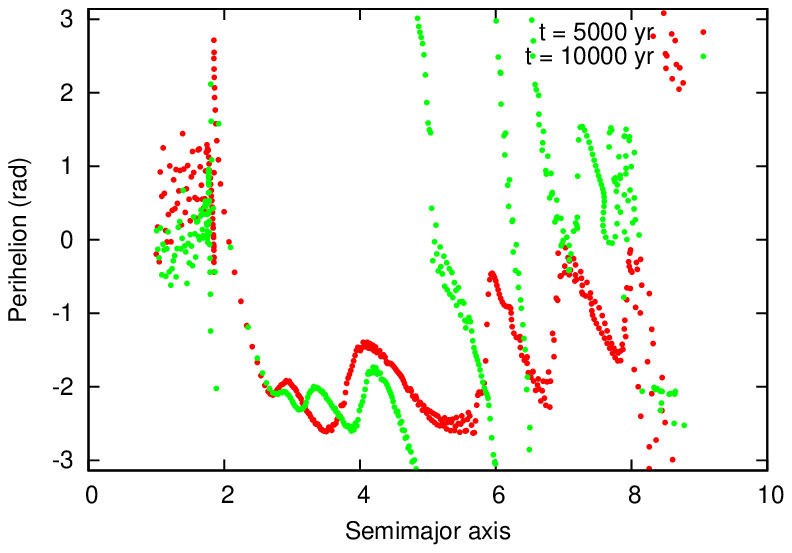}}
\caption{\label{fp1} In panel 1 we show the distribution 
of the eccentricity vs. semimajor axis of 5 km size (radius) 
planetesimals
at two different times. In panel 2 the distribution is that 
of the perihelion longitude vs. semimajor axis.
     }
\end{figure}

In Fig.~\ref{fp1} we illustrate the distribution of the 
eccentricity and perihelion longitude vs. 
semimajor axis for 5 km size (radius) planetesimals. 
A peak in eccentricity with a value around 
0.4 is observed in between 2-3 AU 
with planetesimals quickly drifting inwards. 
The semimajor axis drift rate is in fact strongly 
dependent on the eccentricity of the body in the equation 
given by \cite{adachi76} describing the effect
of the gas drag:

\begin{equation}
\frac{da}{dt}  = -\frac {2} {\tau_{drag}} ( \eta^2 +
\frac {5}{8} e^2 + \frac{1}{2} i^2)^{\frac{1}{2}} \times 
(\eta + \frac{17} {16} e^2 + \frac{1}{8} i^2)
\end{equation}

where 

\begin{equation}
\eta  = \frac {v_{kep} - v_{gas}} {v_{kep}} 
\end{equation}

measures of the amount by which gas orbits the 
star (or star pair)  more slowly than a solid body 
due to the gas's partial pressure support with 
$v_{kep}$ being the local Keplerian velocity. 
The timescale $\tau_{drag}$ is given by:

\begin{equation}
\tau_{drag} = \frac { 8 \rho R} { 3 C_D \rho_{gas} v_{kep}}
\end{equation}

The region developing large eccentricities is then 
depleted on a short timescale. The irregular shape of 
the disk in the inside region is at the origin of 
the large eccentricity values that lead to the 
fast inward drift.  
Farther out, the eccentricity is 
lower but still high compared to the 
pure N--body predictions. In this region the radial 
drift is reduced. The pericenter longitude
is only partly aligned and the phasing depends on time
since it changes with time. Only the region between 
3 and 5 AU seems to maintain some level of coherence 
over time. This coherence leads to lower impact
velocities. 

\subsection{The influence of the disk gravity unveiled}

\begin{figure}[hpt]
\resizebox{90mm}{!}{\includegraphics[angle=0]{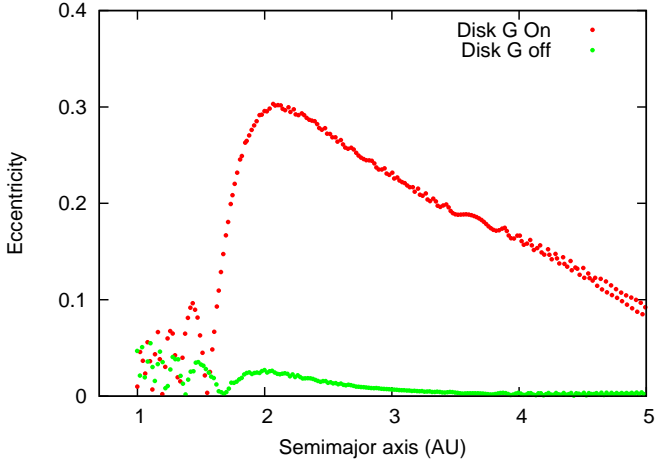}}
\caption{\label{fg1} Test run to verify the impact of the 
disk gravity on the planetesimal dynamics. The outcome of 
the full model is compared with that of a model without 
the gravitational attraction of the disk on planetesimals. 
In this second case the forced eccentricity is significantly 
reduced. The comparison is performed after 1000 yrs of
evolution and for 5 km size planetesimals. 
     }
\end{figure}

By inspecting Fig.~\ref{fp1}, the first question that comes 
to mind is: is it the radial component of the gas drag force 
or the gravitational attraction of the disk that is  
responsible for the large values of the planetesimal 
eccentricities? To answer this question we run an additional 
model where the gravity of the disk on planetesimal 
is switched off. Fig.~\ref{fg1} shows the difference in the 
two cases after $1 \times 10^3$ yrs of evolution. The case
with the gravity of the disk acting on planetesimals show much 
larger eccentricities than the test case without the disk 
gravity. Note that the eccentricity profile in 
the model where the disk perturbations are included 
differ from that shown in Fig.~\ref{fp1} since the 
evolutionary times are different ($10^4$ yr in 
Fig.~\ref{fp1} and $10^3$ yr in Fig.~\ref{fg1}).

\begin{figure}[hpt]
\resizebox{90mm}{!}{\includegraphics[angle=0]{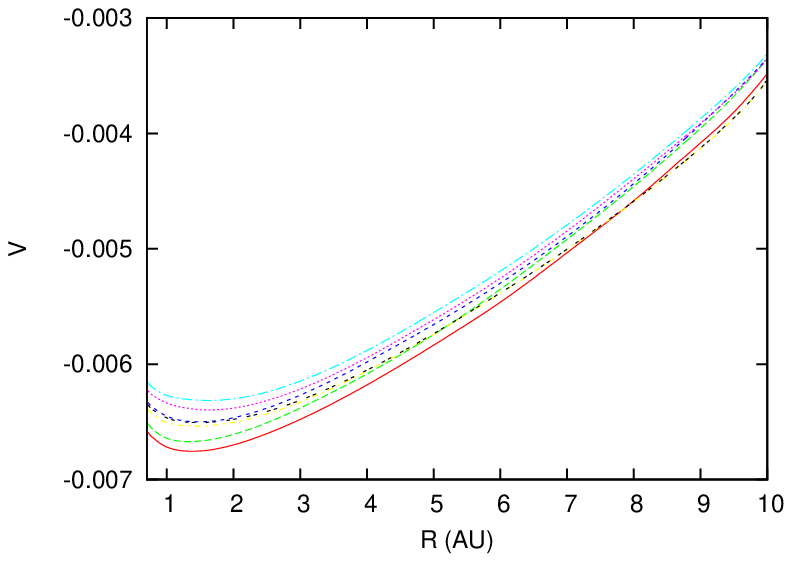}}
\resizebox{90mm}{!}{\includegraphics[angle=0]{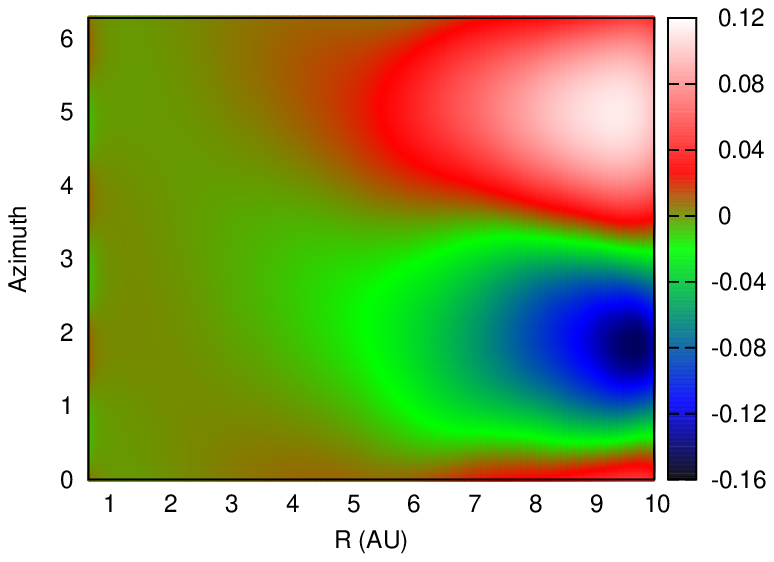}}
\resizebox{90mm}{!}{\includegraphics[angle=0]{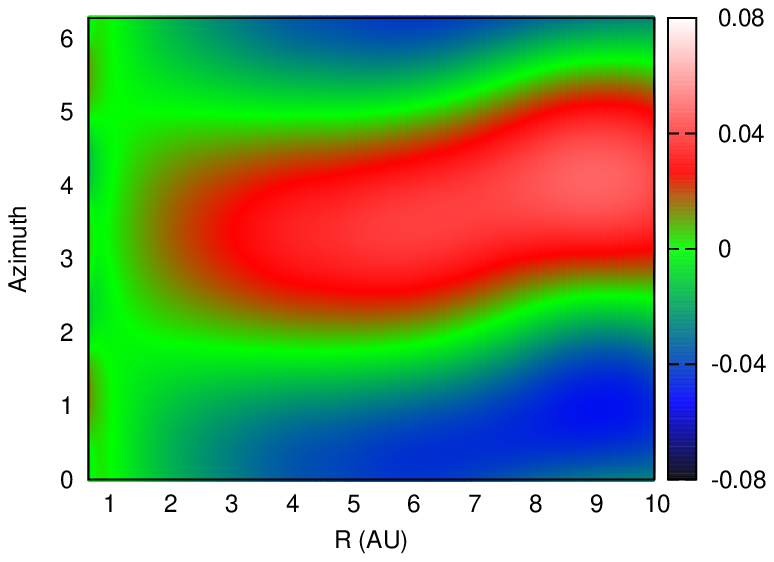}}
\caption{\label{fpot1} Azimuthally averaged 
radial profile of the disk potential (given in normalized
units) 
at different evolutionary
times sampled every 500 yrs. Some change in the potential is also due to the 
mass loss through the inner and outer borders of the 
computational grid. 
The two bottom plots illustrate the azimuthal variation,
computed as $\Delta V = (V - \bar{V}) / \bar{V}$, 
at $t=2500$ and $t=4000$ yr, respectively. The azimuthal variation is not 
regular and depends on the evolutionary time. 
     }
\end{figure}

 While it is possible to analytically estimate the effects 
of a non-linear gas drag due to an eccentric precessing 
gaseous disk \citep{beauge10}, it is a prohibitive 
task trying to predict analytically the details of the 
gravitational perturbation of an asymmetric disk. 
The gravity field felt by each planetesimal is a combination
of the tidal field of the binary stars, of the 
eccentric disk and of its spiral arms which are 
tightly wound close to the stars. Planetesimals are 
well embedded in the disk and then they are sensitive not only to 
the overall shape of the disk but also to its fast 
time variability.  In Fig.~\ref{fpot1} top panel we show 
azimuthally averaged radial profiles of the 
potential 
produced by the disk at different evolutionary times. 
To give an idea of the variation of the potential with 
azimuth we plot also the variation of the potential 
with azimuth respect to the local average value at two
different times. 
The inner zone, within $\sim 1.8$ AU, show a slow 
decrease of the potential with a limited 
azimuthal variability. This is the region where the planetesimals
are less excited in eccentricity. Just beyond 2 AU, 
the potential begins to rise and the dynamics of planetesimals 
reacts to this trend change with a steep increase in eccentricity.
It is possibly the combination of radial and azimuthal variations
that account for the sudden raise in the planetesimal
eccentricity, even if it appears difficult to analytically predict 
the amount of perturbation felt by the planetesimal trajectories. 
This because the whole shape of the potential changes with time. 
In addition, there is no analytical expression available for the 
gravity field of an elliptic disk {\it within} itself. While the 
outside potential might be fitted with the analytical expression 
derived by 
McCoullogh \citep{bookMD}, inside the disk the task appears much 
more complex. 
Not only a secular theory predicting the 
eccentricity perturbations of an elliptical disk 
on bodies located inside the disk itself is intrinsically 
very complex but  to make the task even more difficult the 
disk changes with time and, as a consequence, also the potential,
as shown in Fig.~\ref{fpot1}. 

Even if the effects of the non--radial 
component of gas drag are not relevant in exciting large 
eccentricity values and the disk gravity does 
all the job, this does not mean that the 
gas drag can be neglected when modeling the 
planetesimal evolution. Large values of 
eccentricity powers up 
the gas drag effect on the semimajor axis $a$ since  
$d a / dt$ is proportional to $e$, as discussed in the
previous section. As a consequence, the 
eccentricity excitation leads also to a fast inward 
migration related to the eccentricity value. Does the 
non--radial component of gas drag contributes at later
times to excite the eccentricity to the 
values observed in Fig.~\ref{fp1}--a ? When the 
eccentricity is excited by the disk gravity, the 
large value of $\eta$ is mostly due to the radial 
velocity induced by the eccentric orbit of the planetesimal
rather than the irregular value due to the 
the disk. As a consequence, we expect again that the dominant
term is the disk gravity and that gas drag still 
acts as a damping force. This is further confirmed 
by the case of 25 km size planetesimals discussed in
the next section. It has to be noted that 
the simulations shown in Fig.~\ref{fg1} show the 
eccentricity value after $10^3$ of evolution while those 
in Fig.~\ref{fp1} and Fig.~\ref{fp2} illustrate the 
eccentricity distribution at later times ($10^4$) when 
the planetesimal evolution reaches an almost stationary state.

\subsection{Evolution of 25 km size planetesimals}

\begin{figure}[hpt]
\resizebox{90mm}{!}{\includegraphics[angle=0]{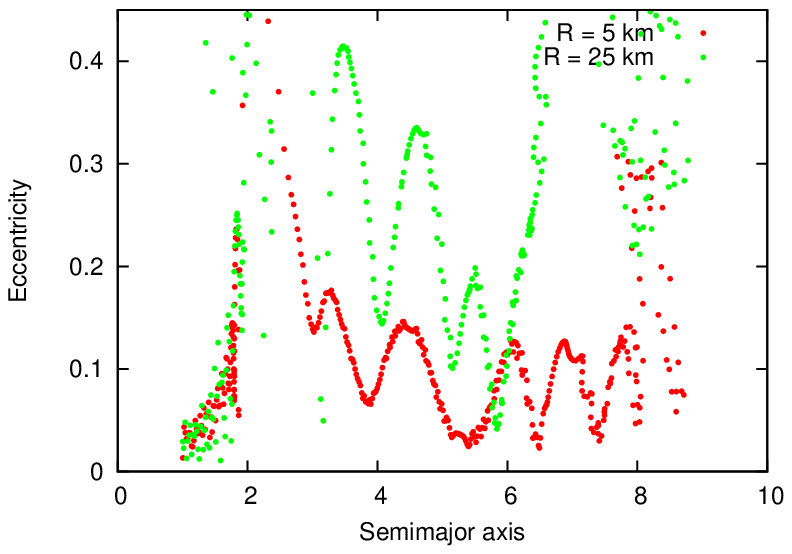}}
\resizebox{90mm}{!}{\includegraphics[angle=0]{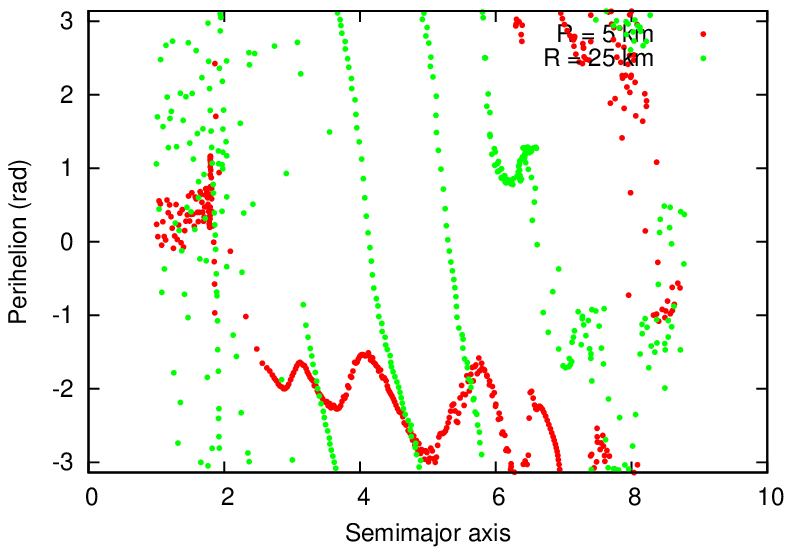}}
\caption{\label{fp2} In panel 1 we compare the distribution
of the eccentricity vs. semimajor axis of 5 km and 25 km size (radius)
planetesimals at $t = 10^4$ yrs. 
In panel 2 the distribution is that
of the perihelion longitude vs. semimajor axis.
     }
\end{figure}

As a further test that the disk gravity is responsible for the 
large eccentricity values of the planetesimals, we ran an additional
simulation for larger bodies 25 km in radius. In Fig.~\ref{fp2}
we compare the eccentricity and pericenter distribution of the 
two different size swarm. The eccentricities are much larger and 
in effect 5\% of the bodies are injected in hyperbolic orbits. 
Their orbits, located in between 2--3 AU where the 
eccentricity excitation is the highest, are perturbed by the disk until they 
reach an eccentricity of about 0.6--0.7 so that, at 
pericenter, they come close to the binary. Repeated pericenter
passages further pump up their eccentricity because of the 
interaction with the binary until a close encounter with
the secondary star 
ejects them out of the system.  
In addition, the pericenters are not phased at all for 
large planetesimals. This is due to the reduced effect of the 
gas drag which was partly able to damp the eccentricities of 
the smaller 5 km size planetesimals but it is unable to perform 
this task for 25 km size bodies. As a consequence, the disk 
gravity is more effective for the large planetesimals in 
exciting their orbits. 

\subsection{Impact velocities}

The dynamical behaviours identified in the previous sections have to 
be interpreted in terms of how they affect mutual accretion of planetesimals. 
The outcome of mutual planetesimal collisions depends on both their 
impact velocities and their respective sizes. Ideally, one would thus 
like to follow a whole population of planetesimals with a given
size distribution and record all collisions for all impacting 
pairs of sizes $R_1$ and $R_2$. Unfortunately, because of the high CPU 
cost of computing planetesimal orbits in this sophisticated set-up, 
only 400 can be followed simultaneously. This is not enough to 
consider a size distribution amongst them and have enough statistics 
on $dv_{R_1,R_2}$ everywhere in the disk. We thus considered 2 simplified 
cases where all planetesimals have the same size, one for "small" 
planetesimals with $R=5\,$km, and one with a "large" planetesimals 
with $R=25\,$km. The consequence of this simplification is to 
underestimate impact velocities among
 planetesimals, as gas drag tends to minimize these velocities 
for equal-sized impactors while increasing them for 
differentially-sized objects \citep{theb06}. As such, our estimates 
should be considered as a best case (that is, accretion-friendly) scenario.

To derive an estimate of the mutual impact velocities between 
the "small" and "large" planetesimals orbiting the binary 
from the orbital parameters of the test bodies in our hybrid model, we used a post--processing 
code computing all possible crossings among the test 
planetesimals of our sample. For each pair of orbits, 
the code looks for the crossing location and 
computes the relative velocity from 2--body keplerian 
formulas. In this way we build up a statistical sample of 
possible impact velocities which characterize the planetesimal 
swarm around the binary. 
These velocities are then compared to the critical velocity $v_{crit(R)}$, corresponding, 
for impacts 
between equal-sized bodies of size R, to the limiting value above which impacts result in 
net mass erosion instead of mass accretion. To derive $v_{crit(R)}$, 
we use Eqs.1 and 2 of 
\cite{linste12}, which lead to $v_{crit(5km)}\sim 25$ m s$^{-1}$ for 5 km size planetesimals
and $v_{crit(25km)}\sim 125$ m s$^{-1}$ for 25 km size planetesimals.

\begin{figure}[hpt]
\resizebox{90mm}{!}{\includegraphics[angle=0]{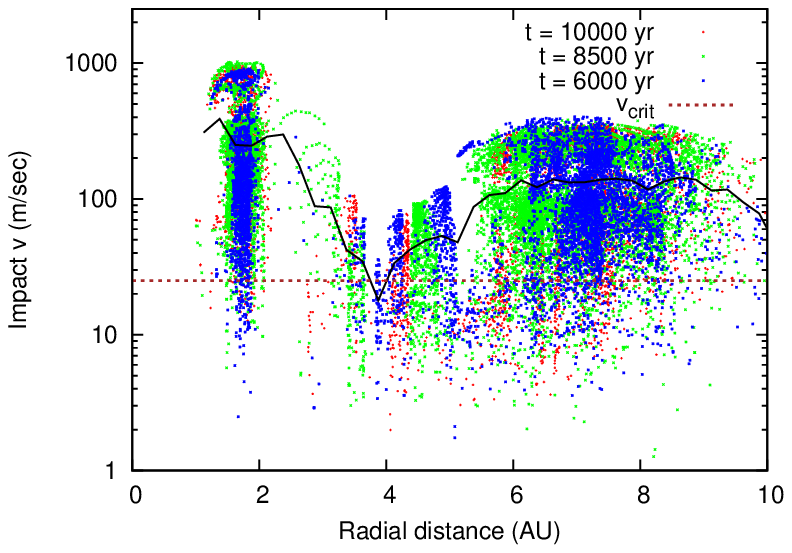}}
\resizebox{90mm}{!}{\includegraphics[angle=0]{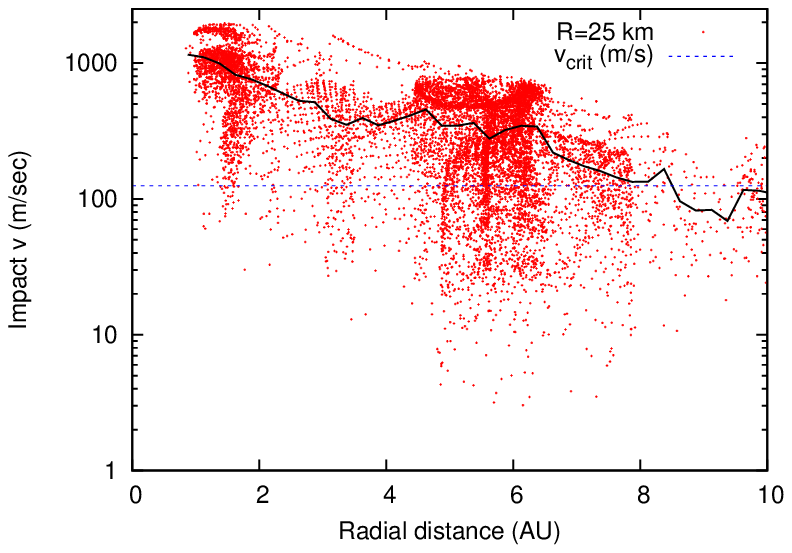}}
\caption{\label{fp3}  Relative encounter velocities between 
planetesimals. In panel 1 we compare the relative velocities 
in the case of small planetesimals at different times. 
There is a significant time variability in the 
impact velocity which, however, remains large all the time. 
In panel 2 we illustrate the relative velocities for 
$R = 25 km$ planetesimals at $t = 10^4 yr$. The black lines 
in both plots show the average impact velocity computed in 
small radial bins
The 2 horizontal lines 
display the limiting $\Delta v$ values for the accretion/erosion frontier, as deduced 
from \cite{linste12}, for $5$\,km impactors and $25$\,km ones.
}
\end{figure}
 
In Fig.~\ref{fp3} we show the relative velocities
among equal size $R = 5 km$ (upper panel) and $R= 25 km$ (lower panel) planetesimals. 
We see that, for both cases, large values of impact speed are excited in the proximity of the 
inner border of the disk, i.e., below $\sim 2\,$AU. This high velocity regime for 
equal-sized bodies is the direct consequence of the gas disk gravity, because pure gas drag would have lead
 to perfect orbital phasing for a given planetesimal size, and thus very low $\Delta v$ for same size bodies.
 However, the effect of gas drag is still noticeable, as it is responsible for the fact that average $\Delta v$
 are higher for $25\,$km planetesimals, $\sim 600-1200$ m s$^{-1}$, than for 5\,km objects, 
$\sim 200-400$ m s$^{-1}$. This difference can be unambiguously attributed to gas drag, as it is the
 only size-dependent mechanism acting on planetesimals, so that its damping effect on the eccentricity will
 be much more pronounced for
smaller bodies. However, even for the small planetesimal run, velocities are still much higher than the 
critical value for erosion $v_{crit(5km)}\sim 25$ m s$^{-1}$. This is also true for 25\,km planetesimals, 
despite of the fact that $v_{crit(5km)}$ is higher, $\sim 125$ m s$^{-1}$. 

The high-$\Delta v$ regime is maintained, for all planetesimal sizes, in the strongly-depleted region between
 2 and 3.5\,AU. It is followed by a region, between $\sim 3.5$ and $\sim 5.5\,$AU, of lower impact velocities
 for the small planetesimal case. However, these velocities remain above the erosion threshold, except
 for a very narrow region around 4 AU, where $<\Delta v>\sim v_{crit}$. For the larger 25\,km planetesimals,
 impact speeds are $\geq 300$ m s$^{-1}$, and thus higher than $v_{crit}$, in this whole intermediate region.

In the outer regions of the simulations, up to 10 AU, impact velocities become comparable for
 both planetesimal populations. Their values remain relatively high, in the $\sim 100-200$ m s$^{-1}$ range,
 and are above the erosion threshold everywhere, except for the outermost 8.5-10\,AU domain,
 where $<\Delta v>$ become slightly lower than $v_{crit}$ for 25\,km objects.

\section{Discussion and Conclusions}
\label{discussion}

The discovery by KEPLER of planets in circumbinary orbits has reawaken the interest
regarding the planetesimal accumulation process in such systems. The first numerical investigations
 of this issue adopted an axisymmetric approximation for the gas disk in which planetesimal are imbedded, 
implicitly assuming that the tidal force of the central binary does not 
significantly perturbs this disk. We show in this paper that this is not the case and, 
due to the presence of the companion star, the circumbinary gas disk becomes eccentric. 
This has profound implications for the accumulation process of planetesimals due to the 
excitation of their orbital eccentricity and partial destruction of the perihelia alignment.
Our simulations where the disk is evolved together with the planetesimals indeed show that, 
adding the crucial effect of the gas disk gravity greatly increases impact velocities amongst 
planetesimals in the circumbinary disc around Kepler 16. 
For the two planetesimal populations we have considered, 5 and 25km, the environment 
is globally very hostile to mutual accretion in the region within $ 10\,$AU. 

As noted earlier, we have considered a simplified case where all bodies have the same size, 
so that only a small fraction of all possible planetesimal encounters have been explored. 
However, in the regions where gas drag has a non-negligible effect on mutual impacting speeds, 
we expect impact velocities between differentially-sized bodies to be even higher 
than those amongst equal-sized ones \citep[e.g.][]{theb06}. As a consequence, in these regions 
our runs probably display a best-case scenario, with a situation that would be even more 
accretion-hostile in a "real" system with a spread in planetesimal sizes. The $r\leq 6\,$AU 
domain clearly corresponds to this case, the important difference between $\Delta v_{5km}$ 
and $\Delta v_{25km}$ being a clear indicator that gas drag is an important factor in imposing 
the impact speeds. In the region beyond 6\,AU, impact speeds for 5km and 25km bodies are comparable, 
which could at first glance indicate that gas drag has a limited effect there. However, 
Fig.\ref{fp2} clearly shows that it is not the case, since these two populations have very different 
eccentricities and periastron in these outer regions, so that gas drag still has an important 
influence on collision velocities and the fact that $\Delta v_{5km} \sim \Delta v_{25km}$ is 
simply a coincidence. We thus here also expect impact speeds to be higher between 
differentially-sized objects than for equal-sized ones, so that our single-sized simulations 
here again probably correspond to the most accretion-friendly configuration possible. The 
fact that even in this best-case scenario the whole disc below $\sim 9\,$AU is strongly hostile to 
accretion, with the possible exception of a narrow strip around $4\,$AU, seems to indicate that 
this result should also hold for any planetesimal population with a spread in its size distributions. 
Furthermore, even if accretion was to be possible, 
it is important to stress that it cannot be as 
efficient as in an unperturbed case. 
Indeed, planetesimal accumulation will be slowed down, 
because of the increased encounter velocities due 
to perturbations which would probably switch off the 
fast runaway-growth mode that requires very low 
impact velocities to proceed 
\citep[see discussion in][]{theb11}.

Our results should, however, be taken with some caution, as one potentially important effect is 
not accounted for in our runs, that of the possible re-accretion of collisional fragments that has 
been identified by Paardekooper et al.(2012). This mechanism would probably help planetesimal growth 
in this dynamically excited environment, so that its omission overestimates the erosive behaviour of the disk. 
Unfortunately, it is impossible to implement this effect in the already very complex set-up that has been considered here.  
We note, however, that other simplifications of our numerical approach might have the opposite effect, i.e., 
underestimating the erosive nature of the planetesimal swarm. The first one is that we suppose that all 
planetesimal start on circular orbits at the same time, whereas there might be a spread in their formation 
epoch \citep[e.g][]{xie10}. Such a spread would lead to increased differential orbits, and thus 
impact speeds, even between equal-sized objects \citep{paar12}. 
The other simplification is that the spatial resolution in the hydrodynamical model required to 
correctly compute the disk--planetesimal interactions limits the radial extension of the disk we can model 
with a reasonable CPU load. Our gas disk is thus probably too small, and we would expect a larger 
and more massive disk to have more powerful gravitational perturbations on the planetesimal 
trajectories, possibly increasing the impact velocities. 

Given these uncertainties, it is thus too early to reach definitive conclusion regarding the 
precise balance between accretion and erosion in a realistic planetesimal disk. However, 
to the very least our simulations have shown that disk gravity plays a crucial role and always 
act towards increasing impact speeds and the erosive behaviour of the swarm. As such, 
they strengthen and expand the results obtained by previous works, which already identified 
that the region where the planet is located is hostile to planetesimal accretion \citep{meschia12,paar12}. 
This seems to rule out the \emph{in situ} formation of the Kepler 16 planet following the 
core-accretion scenario. This result also probably holds for the Kepler-34, Kepler-35 and 
Kepler-47 planets, given the similarities between these different systems.

A possibility to explain the present position of these planets is that they formed 
farther out in the circumbinary disk. Subsequent migration due to the interaction with the disk would have brought them back to their present orbits, as shown in \citep{pinel07, pinel08}. 
This is also suggested by their mass which is in the Neptune--Saturn range, 
in agreement with the prediction of \citep{pinel07, pinel08}. 
Jupiter size planets 
in fact would be either ejected from the system or sent on outer orbits. 
Our simulations show that such migration would have to be very efficient, bringing the 
planet where it is today from an initial formation region probably located beyond 6\,AU from the binary's center of mass.

An alternative scenario may be based on the direct formation of large planetesimals 
from the accumulation of small solid particles in turbulent structures of the 
gaseous disk
\citep{johan07,cuzzi08}. In circumbinary disks the onset of turbulence may be favored by 
the tidal gravity field of the central stars and this might lead to the formation of planetesimal large
enough to sustain the high velocity impacts occurring in the inner regions of the disk. In this case, 
planets might form closer to the center of the disk by--passing the critical phase of small body 
accretion. However, these instability-based scenarios still need to be quantitatively 
investigated in dynamically perturbed environments such as binaries before any conclusion can be reached.

\section*{ACKNOWLEDGMENTS}
We thank an anonymous referee for comments and suggestions. 

\bibliographystyle{aa}
\bibliography{kepler}

\begin{thebibliography}{49}
\expandafter\ifx\csname natexlab\endcsname\relax\def\natexlab#1{#1}\fi

\bibitem[{{Adachi} {et~al.}(1976){Adachi}, {Hayashi}, \& {Nakazawa}}]{adachi76}
{Adachi}, I., {Hayashi}, C., \& {Nakazawa}, K. 1976, Progress of Theoretical
  Physics, 56, 1756

\bibitem[{{Baruteau} \& {Masset}(2008)}]{barmas08}
{Baruteau}, C. \& {Masset}, F. 2008, \apj, 672, 1054

\bibitem[{{Beaug{\'e}} {et~al.}(2010){Beaug{\'e}}, {Leiva}, {Haghighipour}, \&
  {Otto}}]{beauge10}
{Beaug{\'e}}, C., {Leiva}, A.~M., {Haghighipour}, N., \& {Otto}, J.~C. 2010,
  \mnras, 408, 503

\bibitem[{{Bell} \& {Lin}(1994)}]{beli}
{Bell}, K.~R. \& {Lin}, D.~N.~C. 1994, \apj, 427, 987

\bibitem[{{Bitsch} {et~al.}(2013){Bitsch}, {Crida}, {Morbidelli}, {Kley}, \&
  {Dobbs-Dixon}}]{bit13}
{Bitsch}, B., {Crida}, A., {Morbidelli}, A., {Kley}, W., \& {Dobbs-Dixon}, I.
  2013, \aap, 549, A124

\bibitem[{{Boley} {et~al.}(2005){Boley}, {Durisen}, \& {Pickett}}]{Boley05}
{Boley}, A.~C., {Durisen}, R.~H., \& {Pickett}, M.~K. 2005, in Astronomical
  Society of the Pacific Conference Series, Vol. 341, Chondrites and the
  Protoplanetary Disk, ed. A.~N. {Krot}, E.~R.~D. {Scott}, \& B.~{Reipurth},
  839

\bibitem[{{Cuzzi} {et~al.}(2008){Cuzzi}, {Hogan}, \& {Shariff}}]{cuzzi08}
{Cuzzi}, J.~N., {Hogan}, R.~C., \& {Shariff}, K. 2008, \apj, 687, 1432

\bibitem[{{Doyle} {et~al.}(2011){Doyle}, {Carter}, {Fabrycky}, {Slawson},
  {Howell}, {Winn}, {Orosz}, {Prsa}, {Welsh}, {Quinn}, {Latham}, {Torres},
  {Buchhave}, {Marcy}, {Fortney}, {Shporer}, {Ford}, {Lissauer}, {Ragozzine},
  {Rucker}, {Batalha}, {Jenkins}, {Borucki}, {Koch}, {Middour}, {Hall},
  {McCauliff}, {Fanelli}, {Quintana}, {Holman}, {Caldwell}, {Still},
  {Stefanik}, {Brown}, {Esquerdo}, {Tang}, {Furesz}, {Geary}, {Berlind},
  {Calkins}, {Short}, {Steffen}, {Sasselov}, {Dunham}, {Cochran}, {Boss},
  {Haas}, {Buzasi}, \& {Fischer}}]{doyle11}
{Doyle}, L.~R., {Carter}, J.~A., {Fabrycky}, D.~C., {et~al.} 2011, Science,
  333, 1602

\bibitem[{{Duch{\^e}ne}(2010)}]{Duchene10}
{Duch{\^e}ne}, G. 2010, \apjl, 709, L114

\bibitem[{{G{\"u}nther} \& {Kley}(2002)}]{gukle02}
{G{\"u}nther}, R. \& {Kley}, W. 2002, \aap, 387, 550

\bibitem[{{Holman} \& {Wiegert}(1999)}]{holwi}
{Holman}, M.~J. \& {Wiegert}, P.~A. 1999, \aj, 117, 621

\bibitem[{{Hubeny}(1990)}]{Hubeny90}
{Hubeny}, I. 1990, \apj, 351, 632

\bibitem[{{Johansen} {et~al.}(2007){Johansen}, {Oishi}, {Mac Low}, {Klahr},
  {Henning}, \& {Youdin}}]{johan07}
{Johansen}, A., {Oishi}, J.~S., {Mac Low}, M.-M., {et~al.} 2007, \nat, 448,
  1022

\bibitem[{{Kley} \& {Nelson}(2008)}]{klene08}
{Kley}, W. \& {Nelson}, R.~P. 2008, \aap, 486, 617

\bibitem[{{Kley} {et~al.}(2008){Kley}, {Papaloizou}, \& {Ogilvie}}]{kle08}
{Kley}, W., {Papaloizou}, J.~C.~B., \& {Ogilvie}, G.~I. 2008, \aap, 487, 671

\bibitem[{{Leinhardt} \& {Stewart}(2012)}]{linste12}
{Leinhardt}, Z.~M. \& {Stewart}, S.~T. 2012, \apj, 745, 79

\bibitem[{{Lubow} \& {Ogilvie}(1998)}]{Lubow98}
{Lubow}, S.~H. \& {Ogilvie}, G.~I. 1998, \apj, 504, 983

\bibitem[{{Marzari} {et~al.}(2012){Marzari}, {Baruteau}, {Scholl}, \&
  {Thebault}}]{marba2}
{Marzari}, F., {Baruteau}, C., {Scholl}, H., \& {Thebault}, P. 2012, \aap, 539,
  A98

\bibitem[{{Marzari} \& {Scholl}(2000)}]{mascho}
{Marzari}, F. \& {Scholl}, H. 2000, \apj, 543, 328

\bibitem[{{Marzari} {et~al.}(2009{\natexlab{a}}){Marzari}, {Scholl},
  {Th{\'e}bault}, \& {Baruteau}}]{mabash}
{Marzari}, F., {Scholl}, H., {Th{\'e}bault}, P., \& {Baruteau}, C.
  2009{\natexlab{a}}, \aap, 508, 1493

\bibitem[{{Marzari} {et~al.}(2009{\natexlab{b}}){Marzari}, {Scholl},
  {Th{\'e}bault}, \& {Baruteau}}]{marba1}
{Marzari}, F., {Scholl}, H., {Th{\'e}bault}, P., \& {Baruteau}, C.
  2009{\natexlab{b}}, \aap, 508, 1493

\bibitem[{{Marzari} {et~al.}(2008){Marzari}, {Th{\'e}bault}, \&
  {Scholl}}]{marza09}
{Marzari}, F., {Th{\'e}bault}, P., \& {Scholl}, H. 2008, \apj, 681, 1599

\bibitem[{{Masset}(2000)}]{fargo2}
{Masset}, F.~S. 2000, in Astronomical Society of the Pacific Conference Series,
  Vol. 219, Disks, Planetesimals, and Planets, ed. G.~{Garz{\'o}n}, C.~{Eiroa},
  D.~{de Winter}, \& T.~J. {Mahoney}, 75--+

\bibitem[{{Meschiari}(2012)}]{meschia12}
{Meschiari}, S. 2012, \apj, 752, 71

\bibitem[{{Moriwaki} \& {Nakagawa}(2004)}]{morna04}
{Moriwaki}, K. \& {Nakagawa}, Y. 2004, \apj, 609, 1065

\bibitem[{{M{\"u}ller} \& {Kley}(2012)}]{kle12}
{M{\"u}ller}, T.~W.~A. \& {Kley}, W. 2012, \aap, 539, A18

\bibitem[{{Murray} \& {Dermott}(1999)}]{bookMD}
{Murray}, C.~D. \& {Dermott}, S.~F. 1999, {Solar system dynamics}

\bibitem[{{Nelson} \& {Gressel}(2010)}]{grenel10}
{Nelson}, R.~P. \& {Gressel}, O. 2010, \mnras, 409, 639

\bibitem[{{Orosz} {et~al.}(2012){Orosz}, {Welsh}, {Carter}, {Fabrycky},
  {Cochran}, {Endl}, {Ford}, {Haghighipour}, {MacQueen}, {Mazeh},
  {Sanchis-Ojeda}, {Short}, {Torres}, {Agol}, {Buchhave}, {Doyle}, {Isaacson},
  {Lissauer}, {Marcy}, {Shporer}, {Windmiller}, {Barclay}, {Boss}, {Clarke},
  {Fortney}, {Geary}, {Holman}, {Huber}, {Jenkins}, {Kinemuchi}, {Kruse},
  {Ragozzine}, {Sasselov}, {Still}, {Tenenbaum}, {Uddin}, {Winn}, {Koch}, \&
  {Borucki}}]{orosz12}
{Orosz}, J.~A., {Welsh}, W.~F., {Carter}, J.~A., {et~al.} 2012, Science, 337,
  1511

\bibitem[{{Paardekooper} {et~al.}(2011){Paardekooper}, {Baruteau}, \&
  {Kley}}]{pbk11}
{Paardekooper}, S., {Baruteau}, C., \& {Kley}, W. 2011, \mnras, 410, 293

\bibitem[{{Paardekooper} \& {Leinhardt}(2010)}]{Parli10}
{Paardekooper}, S.-J. \& {Leinhardt}, Z.~M. 2010, \mnras, 403, L64

\bibitem[{{Paardekooper} {et~al.}(2012){Paardekooper}, {Leinhardt},
  {Th{\'e}bault}, \& {Baruteau}}]{paar12}
{Paardekooper}, S.-J., {Leinhardt}, Z.~M., {Th{\'e}bault}, P., \& {Baruteau},
  C. 2012, \apjl, 754, L16

\bibitem[{{Paardekooper} {et~al.}(2008){Paardekooper}, {Th{\'e}bault}, \&
  {Mellema}}]{paard08}
{Paardekooper}, S.-J., {Th{\'e}bault}, P., \& {Mellema}, G. 2008, \mnras, 386,
  973

\bibitem[{{Payne} {et~al.}(2009){Payne}, {Wyatt}, \& {Th{\'e}bault}}]{Payne09}
{Payne}, M.~J., {Wyatt}, M.~C., \& {Th{\'e}bault}, P. 2009, \mnras, 400, 1936

\bibitem[{{Pelupessy} \& {Zwart}(2013)}]{pelu13}
{Pelupessy}, F.~I. \& {Zwart}, S.~P. 2013, \mnras, 429, 895

\bibitem[{{Pierens} \& {Nelson}(2007)}]{pinel07}
{Pierens}, A. \& {Nelson}, R.~P. 2007, \aap, 472, 993

\bibitem[{{Pierens} \& {Nelson}(2008)}]{pinel08}
{Pierens}, A. \& {Nelson}, R.~P. 2008, \aap, 483, 633

\bibitem[{{Rafikov}(2012)}]{rafi13}
{Rafikov}, R.~R. 2012, ArXiv e-prints

\bibitem[{{Shakura} \& {Sunyaev}(1973)}]{shak}
{Shakura}, N.~I. \& {Sunyaev}, R.~A. 1973, \aap, 24, 337

\bibitem[{{Thebault}(2011)}]{theb11}
{Thebault}, P. 2011, Celestial Mechanics and Dynamical Astronomy, 111, 29

\bibitem[{{Th{\'e}bault} {et~al.}(2006){Th{\'e}bault}, {Marzari}, \&
  {Scholl}}]{theb06}
{Th{\'e}bault}, P., {Marzari}, F., \& {Scholl}, H. 2006, \icarus, 183, 193

\bibitem[{{Th{\'e}bault} {et~al.}(2008){Th{\'e}bault}, {Marzari}, \&
  {Scholl}}]{thems08}
{Th{\'e}bault}, P., {Marzari}, F., \& {Scholl}, H. 2008, \mnras, 388, 1528

\bibitem[{{Th{\'e}bault} {et~al.}(2009{\natexlab{a}}){Th{\'e}bault}, {Marzari},
  \& {Scholl}}]{thems09}
{Th{\'e}bault}, P., {Marzari}, F., \& {Scholl}, H. 2009{\natexlab{a}}, \mnras,
  393, L21

\bibitem[{{Th{\'e}bault} {et~al.}(2009{\natexlab{b}}){Th{\'e}bault}, {Marzari},
  \& {Scholl}}]{theb09}
{Th{\'e}bault}, P., {Marzari}, F., \& {Scholl}, H. 2009{\natexlab{b}}, \mnras,
  393, L21

\bibitem[{{Th{\'e}bault} {et~al.}(2004){Th{\'e}bault}, {Marzari}, {Scholl},
  {Turrini}, \& {Barbieri}}]{thebs04}
{Th{\'e}bault}, P., {Marzari}, F., {Scholl}, H., {Turrini}, D., \& {Barbieri},
  M. 2004, \aap, 427, 1097

\bibitem[{{Welsh} {et~al.}(2012){Welsh}, {Orosz}, {Carter}, {Fabrycky}, {Ford},
  {Lissauer}, {Pr{\v s}a}, {Quinn}, {Ragozzine}, {Short}, {Torres}, {Winn},
  {Doyle}, {Barclay}, {Batalha}, {Bloemen}, {Brugamyer}, {Buchhave},
  {Caldwell}, {Caldwell}, {Christiansen}, {Ciardi}, {Cochran}, {Endl},
  {Fortney}, {Gautier}, {Gilliland}, {Haas}, {Hall}, {Holman}, {Howard},
  {Howell}, {Isaacson}, {Jenkins}, {Klaus}, {Latham}, {Li}, {Marcy}, {Mazeh},
  {Quintana}, {Robertson}, {Shporer}, {Steffen}, {Windmiller}, {Koch}, \&
  {Borucki}}]{welsh12}
{Welsh}, W.~F., {Orosz}, J.~A., {Carter}, J.~A., {et~al.} 2012, \nat, 481, 475

\bibitem[{{Xie} {et~al.}(2010{\natexlab{a}}){Xie}, {Payne}, {Th{\'e}bault},
  {Zhou}, \& {Ge}}]{xie10b}
{Xie}, J.-W., {Payne}, M.~J., {Th{\'e}bault}, P., {Zhou}, J.-L., \& {Ge}, J.
  2010{\natexlab{a}}, \apj, 724, 1153

\bibitem[{{Xie} \& {Zhou}(2009)}]{xie09}
{Xie}, J.-W. \& {Zhou}, J.-L. 2009, \apj, 698, 2066

\bibitem[{{Xie} {et~al.}(2010{\natexlab{b}}){Xie}, {Zhou}, \& {Ge}}]{xie10}
{Xie}, J.-W., {Zhou}, J.-L., \& {Ge}, J. 2010{\natexlab{b}}, \apj, 708, 1566

\end{thebibliography}

 \end{document}